\documentclass[nonacm]{acmart}

\usepackage{booktabs}
\usepackage{array}

\makeatletter
\let\@authorsaddresses\@empty
\makeatother

\newcommand\blfootnote[1]{%
  \begingroup
  \renewcommand\thefootnote{}\footnote{#1}%
  \addtocounter{footnote}{-1}%
  \endgroup
}

\begin{document}

\title[Who Audits the Auditors?]{Who Audits the Auditors? Recommendations from a field scan of the algorithmic auditing ecosystem}

\author{Sasha Costanza-Chock}
\affiliation{%
  \institution{Northeastern University}
  \country{USA}}
\orcid{0000-0002-0253-9654}
\email{schock@cyber.law.harvard.edu}

\author{Emma Harvey}
\affiliation{%
  \institution{Cornell University}
  \country{USA}}
\orcid{0000-0001-8453-4963}
\email{evh29@cornell.edu}

\author{Inioluwa Deborah Raji}
\affiliation{%
  \institution{University of California, Berkeley}
  \country{USA}}
\email{rajiinio@berkeley.edu}

\author{Martha Czernuszenko}
\affiliation{%
  \institution{Independent Researcher}
  \country{USA}}
\email{mvczer@gmail.com}

\author{Joy Buolamwini}
\affiliation{%
  \institution{Algorithmic Justice League}
  \country{USA}}
\email{joyab@mit.edu}

\renewcommand{\shortauthors}{Costanza-Chock, Harvey, Raji, Czernuszenko, and Buolamwini}

\begin{abstract}
Algorithmic audits (or `AI audits') are an increasingly popular mechanism for algorithmic accountability; however, they remain poorly defined. Without a clear understanding of audit practices, let alone widely used standards or regulatory guidance, claims that an AI product or system has been audited, whether by first-, second-, or third-party auditors, are difficult to verify and may potentially exacerbate, rather than mitigate, bias and harm. To address this knowledge gap, we provide the first comprehensive field scan of the AI audit ecosystem. We share a catalog of individuals (N=438) and organizations (N=189) who engage in algorithmic audits or whose work is directly relevant to algorithmic audits; conduct an anonymous survey of the group (N=152); and interview industry leaders (N=10). We identify emerging best practices as well as methods and tools that are becoming commonplace, and enumerate common barriers to leveraging algorithmic audits as effective accountability mechanisms. We outline policy recommendations to improve the quality and impact of these audits, and highlight proposals with wide support from algorithmic auditors as well as areas of debate. Our recommendations have implications for lawmakers, regulators, internal company policymakers, and standards-setting bodies, as well as for auditors. They are: 1) require the owners and operators of AI systems to engage in independent algorithmic audits against clearly defined standards; 2) notify individuals when they are subject to algorithmic decision-making systems; 3) mandate disclosure of key components of audit findings for peer review; 4) consider real-world harm in the audit process, including through standardized harm incident reporting and response mechanisms; 5) directly involve the stakeholders most likely to be harmed by AI systems in the algorithmic audit process; and 6) formalize evaluation and, potentially, accreditation of algorithmic auditors.

\blfootnote{\copyright 2022. Copyright held by the authors. This is the authors' version of the work. The version of record was published in \textit{2022 ACM Conference on Fairness, Accountability, and Transparency (FAccT ’22)} and is available at https://doi.org/10.1145/3531146.3533213.}
\end{abstract}



\maketitle

\section{Introduction}

\begin{quote}
    “AI auditing isn’t really a thing at this point .. [b]y and large, that’s sort of an aspirational category.” – Interview with Meredith Whittaker, Faculty Director of the AI Now Institute
\end{quote}

As deployed algorithmic products and systems become more common and their harmful impacts more visible, efforts to audit them have become increasingly mainstream. A variety of individuals and organizations now conduct algorithmic audits of products ranging from hiring recommendation engines to facial recognition models, and the algorithmic audit has emerged as one of the most popular approaches to algorithmic accountability \cite{Wilson2021, GenderShades}. Entities that offer audit services have also proliferated, even as audit processes remain unstandardized and poorly understood. For the purposes of this paper, we use the terms `AI audit' and `algorithmic audit' interchangeably to refer to a process through which an automated decision system (ADS) or algorithmic product, tool, or platform (also referred to here under the umbrella term `AI system') is evaluated. An AI auditor evaluates according to a specific set of criteria and provides findings and recommendations to the auditee, to the public, and/or to another actor, such as to a regulatory agency or as evidence in a legal proceeding. In theory, AI audits can help identify whether algorithmic products and systems meet or fall short of expectations in the areas of bias, effectiveness, transparency, direct impacts on vulnerable communities, security and access, regulatory compliance, data consent, labor practices, and/or energy use \cite{Brown2021}. However, without a clear understanding of existing audit practices, let alone widely used standards or regulatory guidance, claims that an AI system has been audited---whether by first-, second-, or third-parties---are difficult to verify and may potentially exacerbate, rather than mitigate, harm.\footnote{For more in-depth definitions of algorithmic systems, algorithmic bias, and algorithmic harm, see Buolamwini's doctoral work \cite{BuolamwiniPHD}.}

To address the gap in knowledge about current AI audit barriers and best practices, we provide the first comprehensive field scan of the AI audit ecosystem. We share a catalog of individuals (N=438) and organizations (N=189) who engage in AI audits or whose work is directly relevant to AI audits, conduct an anonymous survey of the group (N=152), and interview industry leaders (N=10). We identify emerging best practices as well as methods and tools that are becoming widely used, and enumerate key barriers to making AI audits more effective mechanisms for algorithmic accountability. We conclude by outlining six policy recommendations to enable more effective and impactful audit practices. We focus in particular on areas of emerging consensus among auditors, although we also highlight several areas of ongoing debate. Our policy recommendations have implications for the work of lawmakers and regulatory bodies as they develop algorithmic accountability law and regulatory mechanisms, for those responsible for setting and implementing company policies, for standards-setting bodies and professional organizations, and for auditors. Our recommendations include: 1) mandatory, independent AI audits against clearly defined standards, applicable to both AI product owners and operators; 2) required notification of individuals when they are subject to algorithmic decision-making systems; 3) mandated disclosure of key components of audit findings for peer review; 4) consideration of real-world harm in the audit process, including standardized harm incident reporting and response mechanisms; 5) stakeholder participation in audits, in particular by communities most likely to experience harm from the product or tool that is being audited; and 6) a formal system for evaluation and, potentially, accreditation of AI auditors.
\section{Background}
A growing body of evidence demonstrates that algorithmic systems can propagate racism, classism, sexism, ableism, and other intersecting forms of discrimination that cause real-world harm. For example, such systems have been used to wrongfully deny welfare benefits, kidney transplants, and mortgages to individuals of color as compared to their white counterparts, and have contributed to wrongful arrests due to biases in facial recognition technologies, among many other documented instances of harm \cite{AutomatingInequality, WiredBlackKidney, MarkupMortgage, DFPFRT}. Against this backdrop, scholars and practitioners have suggested auditing as one potential method to improve algorithmic accountability \cite{BuolamwiniPHD, Vecchione2021}. Auditing is imagined as a way to explicitly gather and expose evidence of how deployments fall short of performance claims \cite{Diakopoulos2016}. Researchers have also identified how the history, principles, design, and trends of audit studies originating in the social sciences might inform the relatively nascent field of algorithmic auditing. Some have drawn from traditional audit studies to inspire algorithmic audit design methods, while recognizing that there are distinct challenges \cite{Vecchione2021, Sandvig2014}. Recently, AI auditing has become more common in practice, with a growing number of audit reports published in peer-reviewed journals and conference proceedings, documented and self-published by practitioners, or conducted privately by firms \cite{Yee2021, COMPAS, Wilson2021}. As algorithmic auditing proliferates, some researchers have published articles analyzing audit impacts, while others have published systematic literature reviews to categorize audits based on methods and outcomes, as well as to highlight bright spots and identify gaps \cite{Raji2019, Bandy2021, Dillahunt2017}.

AI audits may be classified as first-, second-, or third-party. First-party AI audits are conducted by internal teams, second-party audits are conducted by contractors, and third-party audits are conducted by independent researchers or entities with no contractual relationship to the audit target \cite{UNESCO}. First-party audits are now common at large tech companies, many of which have established auditing teams within their engineering departments. Examples include Facebook's Responsible AI team; Microsoft’s Fairness, Accountability, Transparency, and Ethics (FATE) group; Snap's Inclusive Camera project; Twitter’s Machine Learning Ethics, Transparency, and Accountability (META) team; and Google’s Ethical AI and Responsible Innovation group, as well as many others \cite{FB, MSFT, SnapDEI, TWTR, Google}. In theory, first-party auditors have high levels of access to the algorithms they seek to assess and are well-positioned to ensure that the problems they encounter are addressed; however, the results of first-party audits are not typically disclosed to the public and there is little transparency about whether their recommendations are implemented. There are occasional exceptions: for example, Amazon made headlines in 2018 when it abandoned a potential hiring algorithm after an internal review found that it discriminated against female applicants \cite{ReutersAmazon}. 

An ecosystem of second-party AI auditors is also quickly developing, with some companies specializing in bias audits (for example, O'Neil Risk Consulting \& Algorithmic Auditing, or ORCAA) as well as teams within larger companies (such as Google and IBM) offering reviews of other firms' AI products. Wilson and Mislove, for example, published an audit they performed as contractors for pymetrics, a company that provides candidate screening services. They examined pymetrics' code and tested its performance on deliberately biased data \cite{Wilson2021}. ORCAA audited HireVue, another candidate screening service, using a different methodology, including an assessment of HireVue's documentation and a set of interviews with internal and external stakeholders \cite{HireVueBrookings}. pymetrics' audit, including full methodology and results, was presented at FAccT \cite{PymetricsMITTech}, although this generated some controversy: in an open letter, many FAccT community members criticized pymetrics marketing of a peer-reviewed `third-party audit,' despite the study's being funded by the company and co-authored by multiple pymetrics employees \cite{OpenLetter}. HireVue, meanwhile, was criticized for misrepresenting the results and placing the final audit report behind an NDA \cite{HireVueBrookings}. While second-party auditors are an important part of the ecosystem, these examples highlight the need for standardized audit methods and disclosure policies.

Third-party audits, conducted by fully independent organizations or individuals with no obligation or contractual relationship to the audit target, have been particularly influential in creating public awareness of algorithmic harms. In a widely-cited example of a third-party audit, the investigative journalism group ProPublica analyzed a recidivism risk scoring system sold by Northpointe; ProPublica demonstrated that the scores falsely flagged Black defendants as likely to recidivate at higher rates than they falsely flagged white defendants \cite{COMPAS}. Teams of investigative journalists such as ProPublica, the Markup, and the Associated Press’ (AP) Tracked project now regularly conduct third-party audits of AI systems, as do civil society groups like the American Civil Liberties Union (ACLU), nonprofits like Upturn, law firms like Foxglove, regulators like the National Institute of Standards and Technology (NIST), and independent academic researchers, amongst others \cite{MarkupNaviance, APGunshot, PrivacySOS, Upturn, Foxglove, NIST, Obermeyer2019}. Some of these groups conduct exclusively third-party audits, while others sometimes collaborate with companies or government agencies to conduct second-party audits. 

Although high-level AI principles are now fairly common, they are challenging to operationalize in practice \cite{Whittlestone2019}. Despite a growing body of work that documents various methods for conducting AI audits, there are few widely adopted standards \cite{BarocasSelbst}. There is ongoing discussion and debate about what audits should entail; what tools (if any) are fit for the purpose; and whether audits, impact assessments, or other evaluation approaches are most likely to provide effective oversight and accountability \cite{Metcalf2021, Ayling2021, WiredAccountable}. There is little consensus yet on the proper way to design and conduct an AI audit \cite{Bandy2021, Sandvig2014}. Many AI audits focus on technical implementation of principles; are primarily quantitative in nature; and do not include relevant stakeholders or consider use context or socioeconomic impacts \cite{Brown2021}. Interviews of ML practitioners on commercial product teams conducted by Holstein et al. \cite{Holstein2019} indicate that tools used to assess bias are rarely designed with an understanding of real-world challenges faced by practitioners, and can be hard to apply to instances of unfairness or bias in practice. There are some exceptions; for example, Krafft et al. \cite{Krafft2021} emphasize the importance of community stakeholder involvement in AI audits, and provide a toolkit to aid in community-led technology audits. Buolamwini \cite{BuolamwiniPHD} argues for the importance of evocative audits, which humanize algorithmic harms and show the impact of systematic bias on specific individuals. Additionally, some organizations have begun to leverage the public to help them identify potential algorithmic harms. For example, after users noticed bias in Twitter's image saliency and cropping algorithm, Twitter launched an algorithmic bias bounty competition at the 2021 DEF CON AI Village, where they made their saliency algorithm and image cropping code publicly available so that participants could win cash prizes for demonstrating bias and potential harm \cite{Yee2021, TwitterBounty}.\footnote{A recent white paper by the Algorithmic Justice League (AJL) provides a deep dive into the requirements for effective algorithmic bias and harm bounty programs \cite{BiasBounty}.} 

AI audit standards remain unclear, but policymakers have begun to draft, introduce, and in some cases pass legislation that requires AI audits, impact assessments, or other forms of evaluation. Legal mechanisms to require and govern audits have longstanding precedent in the financial sector, including through the Securities Act of 1933, the Sarbanes Oxley Act of 2002, and the Dodd-Frank Wall Street Reform and Consumer Protection Act of 2010 \cite{SecAct1933, SarbanesOxley, Dodd}. Some legal AI audit requirements already do exist. At the municipal level, New York City passed regulation in 2021 (taking effect in 2023) that requires mandatory third-party audits of AI hiring and employment systems -- the first of its kind in the US \cite{NYCAudit, CDTNYCAudit}. More comprehensive legislation is pending. In the US, the proposed Algorithmic Accountability Act would include mandatory audits for AI system vendors and operators \cite{AAAct, WiredAccountable}. Two proposed regulations in the EU, the AI Act and the Digital Services Act (DSA), would limit use and facilitate audits of `high-stakes' algorithms and require independent audits of `very large online platforms', respectively \cite{EUAIAct, EUDSA}. However, policy priorities and approaches to the governance of algorithmic systems remain hotly contested. For example, when New York City released its first report on ADS in 2019, it was publicly critiqued by members of the city's own task force, who felt the recommendations were weak and did not reflect consensus \cite{NYCADS}. Similar tensions have bubbled up between civil society and corporate interests, such as when the international NGO Access Now resigned from the Partnership on AI (PAI) in 2020, citing an "increasingly smaller role for civil society to play within PAI” \cite{AccessNow}. There remains a lack of consensus between policymakers and practitioners about what constitutes algorithmic audits, whether they are effective, and even what they should be called. Since different communities of practice hold different priorities and viewpoints, it can be difficult to conceptualize, define, and put meaningful algorithmic audit policies into practice \cite{Krafft2020}. 

Against this backdrop, we set out to identify key policy recommendations for effective AI audits, grounded in practitioner insights, that might be useful for lawmakers and regulators, professional and standards organizations, companies, public interest advocates, and independent researchers alike. To this end, we interviewed and surveyed industry leaders to address five key research questions: 1) what methods and tools are practitioners currently using to audit AI systems?; 2) what are the emerging standards and best practices in AI audits?; 3) what are some of the biggest barriers to effective AI auditing?; 4) do practitioners currently investigate potential and real harm across the AI lifecycle?; and 5) do practitioners specifically pay attention to harm incident reporting for deployed systems?

Our work builds on past studies that have conducted interviews and surveys with AI practitioners. Veale et al. \cite{Veale2018} interviewed public sector ML practitioners, who identified changes in modeling data over time, incorporating human discretion in interpreting or augmenting model outputs, and communicating details on model performance without being misleading, among other dynamics, to be key challenges to incorporating fairness considerations into ML work. More recently, Holstein et al. \cite{Holstein2019} interviewed private-sector machine learning practitioners about challenges to enacting fair AI principles in practice. Their interviewees identified issues including lack of access to necessary data, missing knowledge and biases within AI teams themselves, narrow or overly quantitative approaches to fairness, and lack of proactive auditing processes or commitment to address fairness issues. We find many of these same issues as barriers to conducting AI audits; our research, however, also seeks to understand the methods and tools that are currently used, emerging best practices espoused by industry leaders, and practitioners’ approaches to incorporating analysis of real-world harm and stakeholder engagement into their audit processes. In addition, while Holstein et al. focus on ML teams in the private sector and Veale et al. interview public sector ML practitioners, we focus specifically on algorithmic auditors. We interview and survey first-, second-, and third-party auditors to better understand the similarities and differences between these groups. We also survey civil society advocates, researchers, and regulators to understand areas of agreement and discord between auditors and non-auditors. Finally, while both Veale et al. and Holstein et al. focus primarily on how researchers can overcome the barriers they identify, we also emphasize areas where we believe that government regulation and/or action by professional or standards organizations is necessary.

\begin{table*}
\parbox[t]{.74\linewidth} {
\centering
\caption{Interviewees\label{interview}}
\begin{footnotesize}
\begin{tabular}{m{2cm} m{2.25cm} m{2.5cm} m{2.25cm}}
    \toprule
    \textbf{Name or ID} & \textbf{Role} & \textbf{Organization} & \textbf{Title} \\
    \midrule
    Frida Polli, \newline Alex Vaughan & First-party auditors & pymetrics & CEO, \newline Chief Science Officer \\
    Cathy O'Neil & Second-party auditor & ORCAA & CEO \\
    Liz O'Sullivan & Second-party auditor & Parity & CEO \\
    Reid Blackman & Second-party auditor & Virtue & CEO \\
    Michelle Lee & Second-party auditor & Deloitte UK & AI Ethics Lead \\
    Meredith Whittaker & Researcher & AI Now & Faculty Director \\
    R1 & First-party auditor & Social media company & Internal audit lead \\
    R2 & Third-party auditor & Media organization & Journalist \\
    R3 & First-party auditor & Technology company & Internal audit lead \\
    R4 & First-party auditor & Technology company & Internal audit lead \\
    \bottomrule
\end{tabular}
\end{footnotesize}
}
\hfill
\parbox[t]{.25\linewidth}{
\centering
\caption{Survey respondents \label{survey}}
\begin{footnotesize}
\begin{tabular}{p{2.1cm} p{.9cm}}
    \toprule
    \textbf{Role} & \textbf{Number} \\
    \midrule
    Auditor & 56 \\
    Non-Auditor & 96 \\
    \midrule
    First-party auditor & 12 \\
    Second-party auditor & 16 \\
    Third-party auditor & 11 \\
    Advocate & 65 \\
    Researcher & 96 \\
    Regulator & 3 \\
    Other & 25 \\
    \bottomrule
\end{tabular}
\end{footnotesize}
}
\Description{Two side-by-side tables. The first table outlines the list of AI auditors interviewed, their role (first, second, or third party auditor or researcher), the organization for which they work, and their job title. In terms of first party auditors, we spoke to: Frida Polli and Alex Vaughan, the CEO and CSO of pymetrics, respectively; as well as three anonymous internal audit leads from technology and social media companies. In terms of second-party auditors, we spoke to: Cathy O'Neil, CEO of ORCAA; Liz O'Sullivan, CEO of parity; Reid Blackman, CEO of Virtue; and Michelle Lee, AI Ethics Lead at Deloitte UK. We also spoke to an anonymous third-party auditor and journalist. Finally, we spoke to Meredith Whittaker, a researcher and Faculty Director of AI now. The other table describes our survey respondents. We surveyed 56 auditors and 96 auditors. We surveyed: 12 first-party auditors, 16 second-party auditors, 11 third-party auditors, 65 advocates, 96 researchers, 3 regulators, and 25 individuals who identified as `other' (respondents could select multiple roles).} 
\end{table*}

\section{Methods}

We first conducted a field scan that yielded a list of 438 individuals from 189 organizations that we identified as involved, to some degree, in AI auditing. This list included first-, second-, and third-party auditors as well as members of civil society and advocacy organizations, academic researchers, regulators involved with AI audit legislation and policy, and others involved in audit-related work. From this list, we identified 10 leaders in the field for semi-structured interviews; interviewees are described in Table \ref{interview}.\footnote{All interviewees were offered the option to receive attribution or remain anonymous. Interviewees who wished to receive attribution are identified by name; interviewees who wished to remain anonymous are identified only as R1, R2, R3, and R4. Interviewees were provided with transcripts after the interviews to verify that all quotes were accurately recorded; anonymous interviewees were also given the option to redact potentially identifying quotes. See Appendix \ref{methods-interviews} for more information on interview methods.} Based on our understanding of the field and lessons from our interviews, we then developed a survey that we shared with contacts at all 189 organizations identified in the field scan. Overall, we received 152 individual survey responses. Table \ref{survey} summarizes our survey respondents: 56 (37\%) of respondents indicated that they had personally participated in an audit of an AI system. Of those, 12 were first parties, 16 were second parties, and 11 third parties (respondents could select multiple roles). The auditors had collective experience assessing algorithmic systems in industries including technology and social media, hiring and HR, consumer goods, and insurance and credit, as well as for state and local governments. The full list of identified individuals and organizations, the structured interview guide, and the survey questions can be found in the Appendices. 

\subsection{Limitations}
Several factors potentially limit the accuracy and generalizability of our findings. First, our respondents may not be representative of the entire population of AI auditors. Only 56 survey respondents self-identified as having worked directly on audits. Additionally, a number of AI audits were published after our survey closed; we were not able to invite all of the authors of these new audits to participate in our study. Our survey responses came primarily from respondents in the US (n=90), UK (n=16), and EU (n=18); we only received one response each from auditors in Africa, Asia, and South America, and no responses from auditors in Australia. Our team is based in the US, our materials and outreach were all in English, and we did not systematically attempt to identify and include auditors from all regions of the world. There is a clear need for future research that focuses on algorithmic audits in the Global South. 

Since the very meaning of algorithmic audits is up for debate, our decision to focus on professionalized auditors also influences our findings, as we did not systematically include those working on other approaches to auditing such as community-based, participatory, or evocative audits. As previously mentioned, there is active discussion and debate about different approaches to evaluation of AI systems. For example, some argue that AI impact assessments are a more effective accountability mechanism than audits, or that red teams should be the preferred approach \cite{AINowImpact, Brundage2020}. We sought to include practitioners who use different language and framing in their work to evaluate AI systems, but may have inadvertently excluded some who do not agree with the `audit' framing. Finally, although we firmly believe that the question `who audits?' matters, in this study we did not collect auditors' demographic information. We believe that future work to systematically explore the demographics of algorithmic auditors is necessary.
\section{Key Findings}

\subsection{Methods and Tools Used to Audit AI Systems}

\subsubsection{Quantitative over qualitative}
\begin{quote}
“Any good AI audit has to have some sort of measurable, code-based aspect to it.” – Interview with R1, internal audit lead at a social media company
\end{quote}

\begin{figure*}
\caption{Areas that AI auditors reported assessing. Quantitative assessments are more common than qualitative assessments.
\label{methods}}
\includegraphics[width=\textwidth]{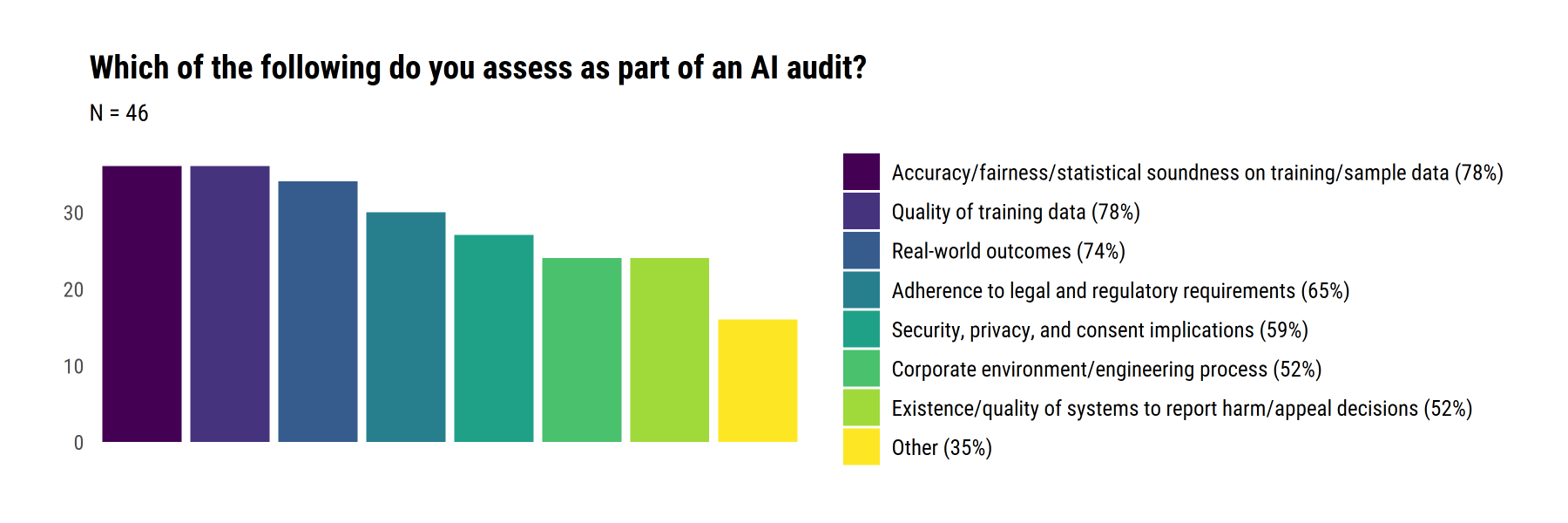}
\Description{A bar graph showing the proportion of surveyed AI auditors (N=46) who reported assessing different components of an AI system as part of an AI audit. `Accuracy/fairness/statistical soundness on training/sample data' and `quality of training data' (both 78\%) are most common, followed by `real-world outcomes' (74\%); `adherence to legal and regulatory requirements' (65\%); `security, privacy, and consent implications' (59\%), and finally `corporate environment/engineering process' and `existence/quality of systems to report harm/appeal decisions' (both 52\%)}
\end{figure*}

AI auditors who completed our survey are more likely to report engaging in quantitative audit methods than critical, structural, or qualitative ones. As depicted in Fig. \ref{methods}, 77\% of auditors say that they assess algorithmic accuracy, fairness, and statistical soundness; 77\% also say they assess the quality of training data. In comparison, just 51\% say they assess the existence and quality of systems for individuals to report real-world harm; similarly, 51\% say they assess the corporate environment and engineering process in which the AI system was created. The four most common methods (over 70\% of respondents employ each) are quantitative: checking whether training data is appropriate for modeling, assessing data representativeness, assessing bias in input data, and measuring accuracy of the AI system on individual subgroups. The four least common methods (under 40\% of respondents employ each) are structural or qualitative: assessing robustness of the AI system to adversarial use, training auditee’s employees on identifying biases and harm, examining team diversity, and assessing whistleblower protection mechanisms.

\subsubsection{Reliance on custom frameworks and tools\label{tools}}
Most practitioners indicate that their auditing frameworks and tools are custom-built, and often tailored to particular use cases. Just 7\% say their audit process uses a standardized framework and set of tools. When asked about specific pre-built AI audit tools such as IBM AI Fairness 360, SciKit Fairness, or Parity (full list available in the survey instrument in Appendix \ref{survey_instrument}), 38\% of respondents say they do not use any. Reid Blackman, CEO of Virtue, described the drawbacks of relying on pre-built tools: "[These tools] use quantitative `definitions' of fairness to assess whether or not the distribution of goods and services across various subpopulations are equitable or fair… You can’t be fair according to all metrics at the same time, and so a substantive ethical qualitative decision needs to be made: ‘which among these various metrics are the appropriate ones to use for this particular use case?’"

\subsubsection{Focus on legally protected classes; intent towards intersectional analysis but thin evidence in practice}
Antidiscrimination law fails to protect many vulnerable sub-populations from harm. According to Liz O’Sullivan, CEO of Parity, “A lot of what we're seeing right now is people are starting with the big legally protected categories. But more than that, companies are coming to us and saying, ‘what haven't we thought of?’” Most auditors (74\%) self-report assessing fairness across not only legally protected demographic categories, but also other categories of interest. However, when asked specifically what protected classes their audits consider, the only categories selected by more than 50\% of auditors are ‘Age,’ ‘Race/Ethnicity,’ and ‘Sex’---three classes that are frequently legally protected. When asked about groups outside of legally protected classes, some auditors describe challenges to including them in audits. For example, when asked about transgender and intersex individuals, Alex Vaughan, Chief Science Officer at pymetrics, states that “our clients are not legally required to assess potential disparate impact against protected classes that are very, very small,” adding that “sample size issues are very real for these few small classes.” Similarly, although a majority of auditors self-report conducting intersectional, as opposed to single-axis, fairness assessments (65\%), we are not able to assess this claim, since very few auditors provided access to examples of audit methods and outcomes. The challenges associated with selecting subgroups for fairness assessment are well-documented by Raji et al. \cite{Raji2020} and Barocas et al. \cite{Barocas2021}, who note that intersectional and/or rare groups are often hard to identify and underrepresented in the data. Respondents also report legal difficulties; Frida Polli, CEO of pymetrics, states “there is the possibility that were you to basically start to de-bias your algorithms based on intersectionality, you might run afoul of disparate treatment law… because it's not a protected class.”

\subsubsection{Limited disclosure and peer review}
\begin{quote}
“[The] best case scenario would be to just require transparency, to have people publish their metrics to an external body, hopefully something in government…[But] the balance between transparency and business is a really tough one to strike.” - Interview with Liz O’Sullivan, CEO of Parity
\end{quote}

Most auditors do not publicly share audit outcomes or documentation of their methods. Of the 43 auditors asked to share a link to their audit process, only seven provide a link to any documentation, and just four link to audit results. Multiple respondents say they cannot share due to client confidentiality agreements. Most (82\%) agree in principle to making audit results, or at least some degree of documentation, publicly available. Publicly releasing audit results does not seem to reduce trust in auditors -- in fact, we found the opposite. When asked to name “best in class” auditors, respondents identified individuals or organizations that tend to publish their audit methodologies and results. All individuals mentioned by multiple respondents have published at least one audit: Cathy O’Neil (mentioned by 9 respondents); Dr. Joy Buolamwini, Dr. Timnit Gebru, and Deborah Raji (6 respondents); Christo Wilson (3 respondents); and Dr. Rumman Chowdhury (2 respondents) \cite{HireVue, GenderShades, Raji2019, Wilson2021, Yee2021}.

\subsection{Emerging Standards and Best Practices in AI Audits}

\subsubsection{Standards are thin, at best}

\begin{quote}
“There are literally no standards […] and there are no best practices”-- Interview with R1, internal audit lead at a social media company 
\end{quote}

Although recent years have seen some progress on AI regulatory oversight and standards, the sentiment expressed by R1 above is shared across interviewees, such as Blackman, (“the standard practice is just to ignore [ethical risk]”), R4, internal audit lead at a technology company, (“I don’t think there are best practices”), and O’Neil (“I have my idea for what best practices would be, but I don’t think they’re happening.”) Survey respondents also overwhelmingly describe regulation related to AI audits as “not or barely present” (75\%); a further 24\% say it “needs to be strengthened.” Just 1\% describe current regulation as “sufficient.” This is generally consistent across geographies, although regulation in the US is considered less developed (82\% say “not/barely” present) than regulation in Europe (56\% in the UK and 67\% in the EU say “not/barely” present). Additionally, 89\% “strongly” or “somewhat” agree that companies do not take action on ethical AI issues unless they face public pressure. R4 corroborates this: “The best way to make sure that the issue is attended to is to have the press report on it,” as does O’Neil: “[Audits] won't happen until there's the right leverage, and that leverage has to either come from regulatory bodies or plaintiff lawsuits or class action lawsuits or reputational risk. And right now, we only have reputational risk. Sometimes.” We also find no single source for best practices, with most respondents indicating they seek information from a variety of avenues, including conferences (FAccT, AIES), professional organization (IEEE, NIST), social media, and civil society (AJL, ForHumanity, Data \& Society).

\subsubsection{Agreement and differences between auditors and non-auditors}
There is a consensus that audits require both technical and qualitative analyses. According to R1, “For an AI audit to work… [it needs to] be interdisciplinary, [including] legal folks, product folks, and model developers.” Of survey respondents, the vast majority consider external (77\%) and internal (79\%) qualitative risk assessments as well as technical review of both inputs (78\%) and outputs (80\%) as necessary features of ‘good’ AI audits. Auditors, however, place slightly more emphasis on technical assessments while non-auditors place slightly more emphasis on qualitative assessments. Additionally, we find that 74\% of all respondents either “somewhat” or “strongly” disagree with the statement that “AI bias audits should focus on evaluating disparate outcomes; it is not important to consider the code and methodology used to create the AI system when determining whether bias is present.” However, among first-party and third-party auditors surveyed, just 33\% and 38\% disagree, respectively. 

\subsubsection{Broad support for regulation requiring audits, notification, and disclosure of key audit results, but disagreement on the details}

\begin{figure*}
\caption{AI auditor consensus and disagreement about regulation.
\label{regulation}}
\centering
\includegraphics[width=0.33\textwidth]{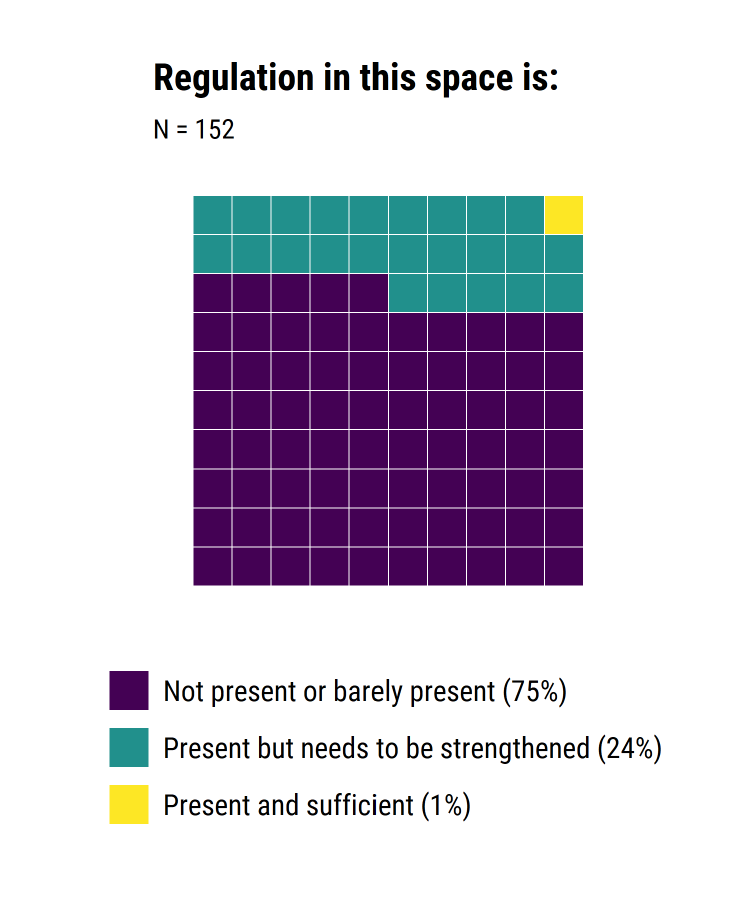}
\includegraphics[width=0.33\textwidth]{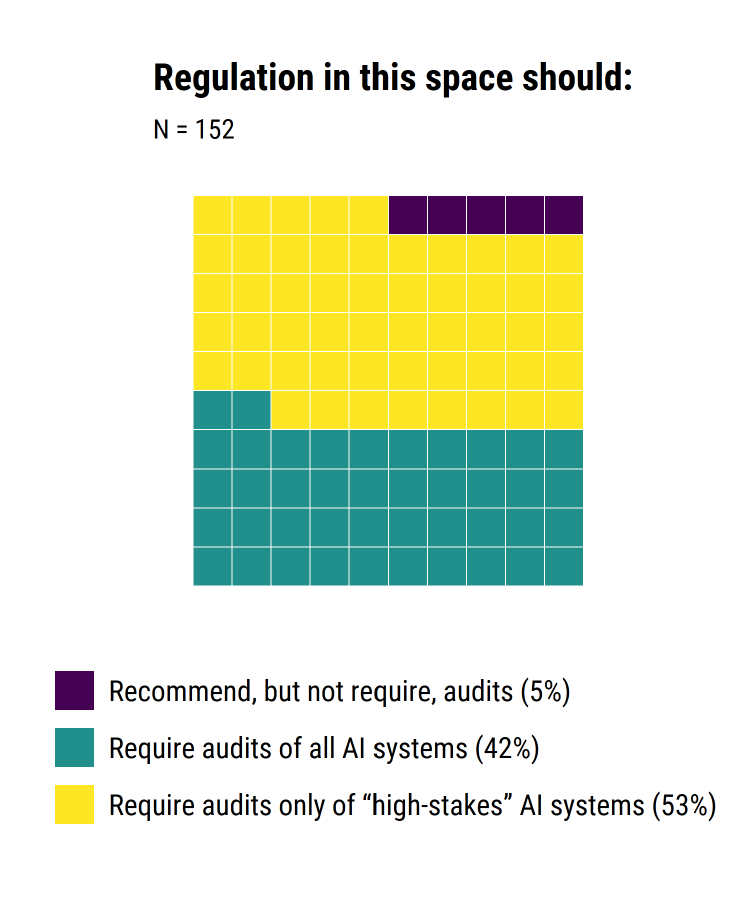}
\includegraphics[width=0.33\textwidth]{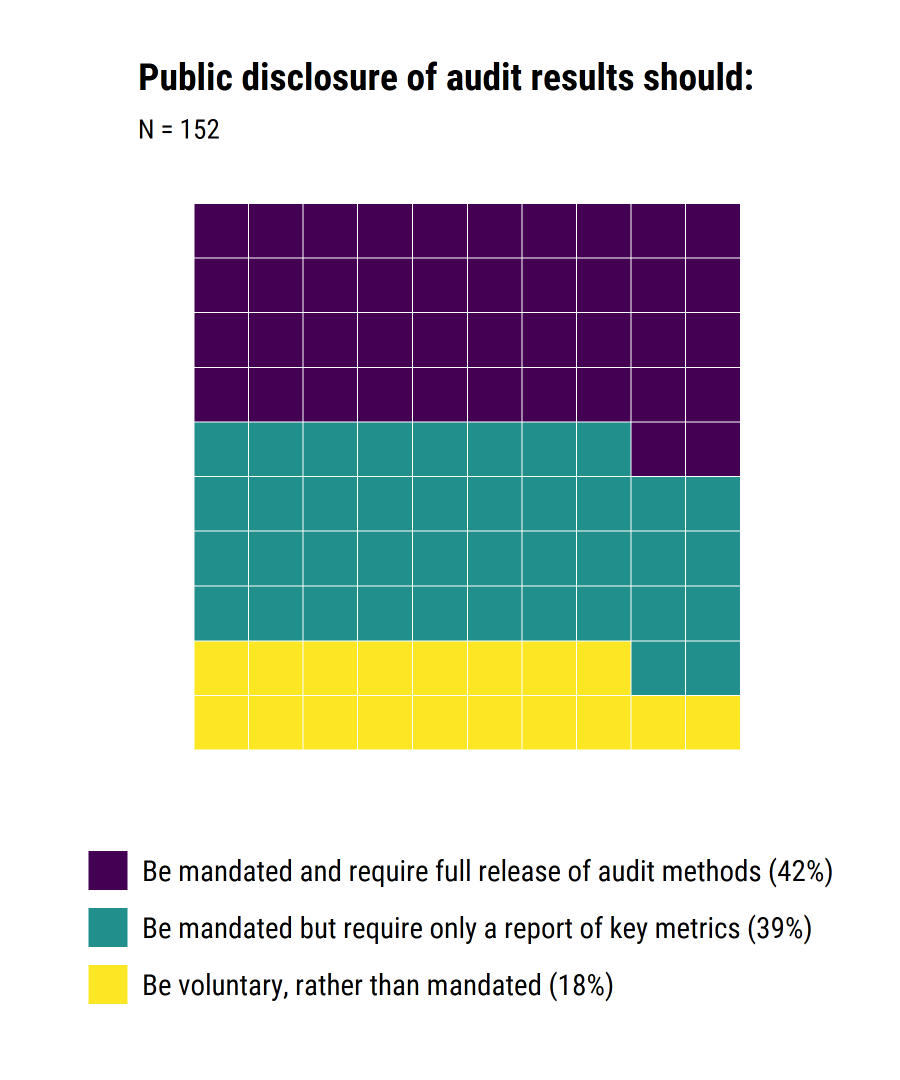}

\includegraphics[width=0.33\textwidth]{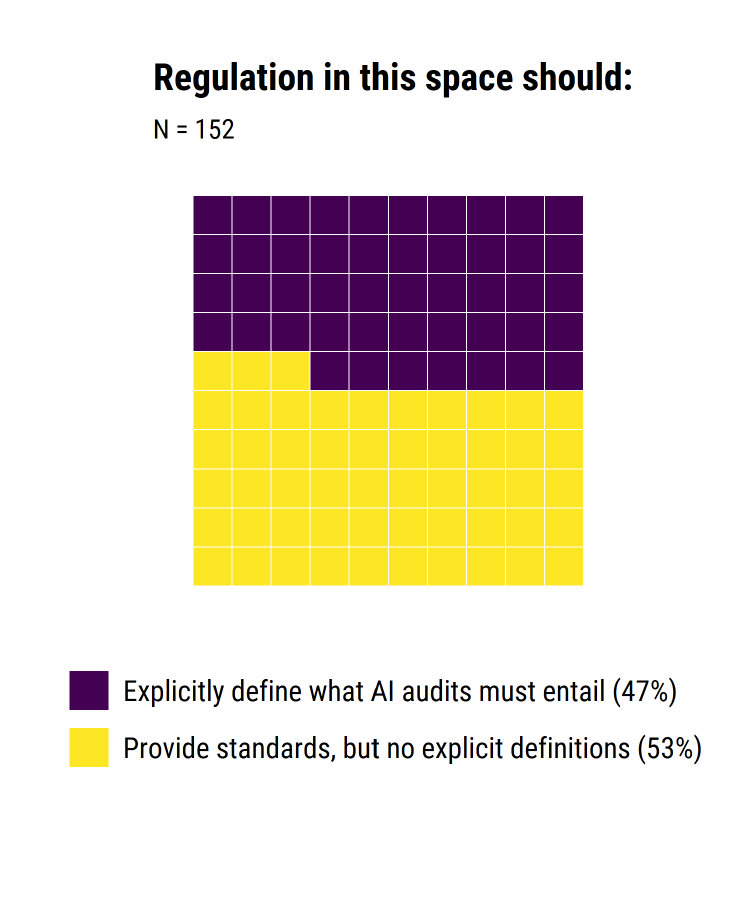}
\includegraphics[width=0.33\textwidth]{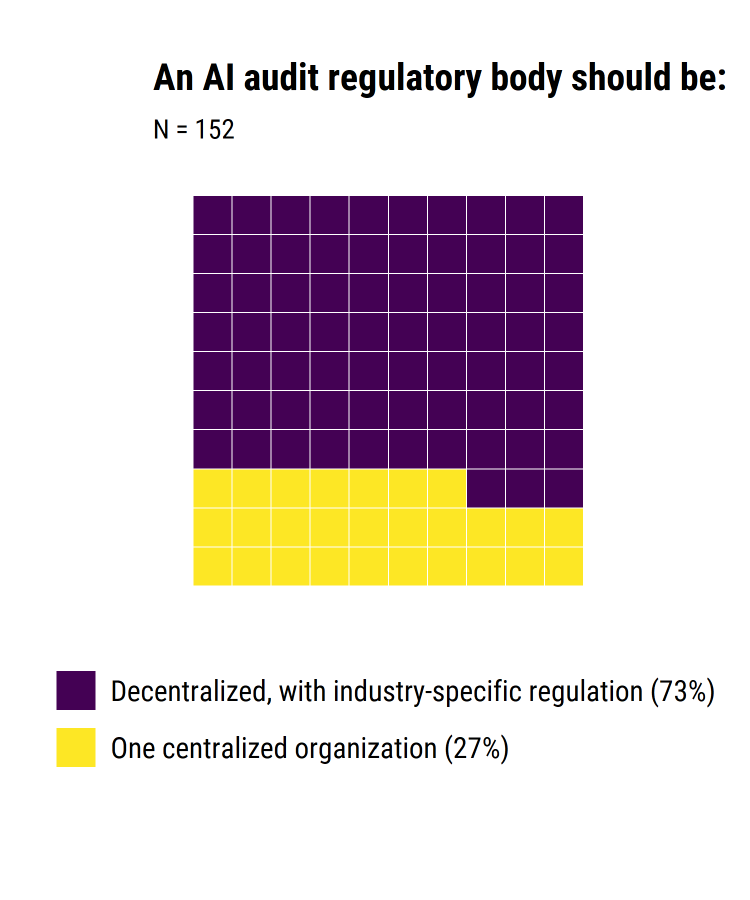}
\includegraphics[width=0.33\textwidth]{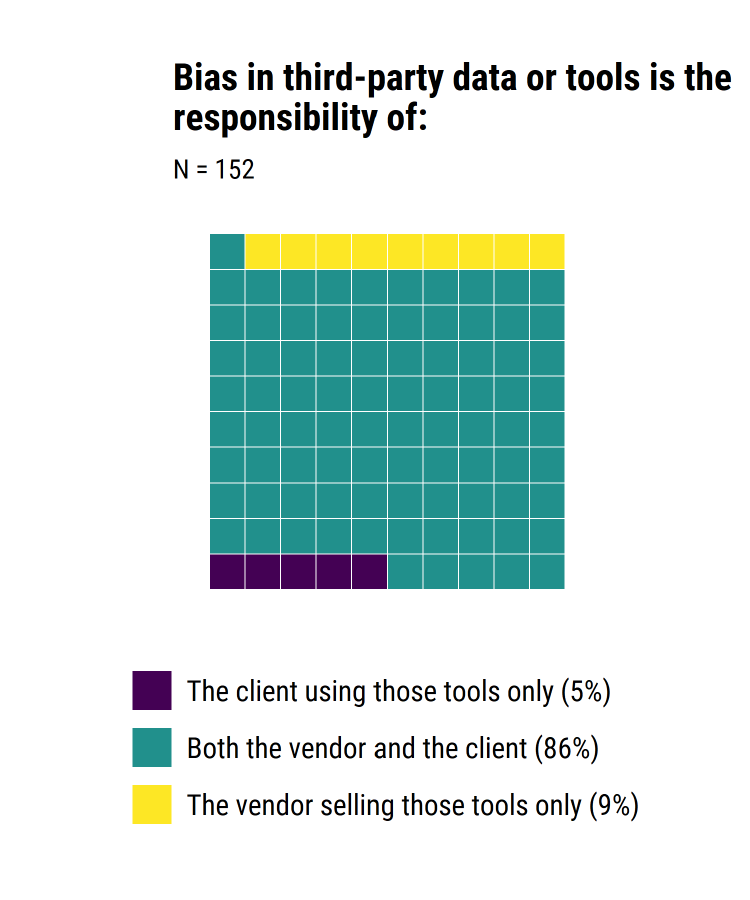}
\Description{Six waffle plots showing agreement and disagreement between our survey respondents (N=152) on a variety of topics. They show: (1) consensus (99\%) that AI audit regulation is either not present or needs to be strengthened; (2) consensus (95\%) that regulation should require AI audits, but disagreement on whether audits should be required for all or only high-stakes systems (42\% and 53\%, respectively); (3) consensus (82\%) that disclosure of audit results should be mandated, but disagreement on whether disclosure should include full results or key metrics only (42\% and 39\%, respectively); (4) disagreement on whether regulation should explicitly define audits (47\%) or only provide standards (53\%); (5) consensus that an AI audit regulatory body should be decentralized and industry specific (73\%); and (6) consensus that bias in third-party tools or data should be the responsibility both of the vendor selling the tools and the client using the tools (86\%).} 
\end{figure*}

We find a clear consensus that the top regulatory priority (among 152 respondents who ranked 13 options) is legislation requiring AI audits. The second-highest ranked priority is legislation requiring that individuals be notified when they are subject to automated decision systems. This was also supported by our interviewees, such as R2, a journalist and third-party auditor: “The use of algorithms is most harmful to people in places where it's the least visible.” Establishment of systematic harm incident reporting was the third-highest ranked regulatory intervention. There are two other notable points of agreement: 73\% believe that a government body regulating AI audits should be decentralized, with multiple agencies regulating domain-specific AI systems. As R2 states, “I believe that domain expertise is so important in auditing algorithms… every field needs to be emboldened with the technology and expertise to analyze them.” Additionally, although there is a tendency for AI vendors to blame clients for improper use, and for clients to blame vendors for flawed models, respondents agree (86\%) that both vendors and clients are responsible for AI harm. Regulation to accredit auditors or certify algorithmic products is less popular. The lowest-ranked regulatory interventions were (1) establishment of a centralized body to oversee AI systems (likely because most auditors indicate a preference for a decentralized regulatory body), (2) formal accreditation of AI auditors, and (3) certification of AI systems as having passed AI audits. 

Although respondents agree that regulation is necessary, they disagree on the details. While there is very strong consensus that audits should be mandated (95\%), respondents disagree on mandate scope (53\% support mandates for ‘high-stakes’ systems only; the other 42\% support mandates for all AI systems). Similarly, of the 82\% who believe that disclosure of audit results should be mandated, 39\% say that disclosure should include only a read-out of key metrics while 42\% believe that the full audit code, methodology, and outcomes should be released. Respondents are also almost perfectly split between those who say future regulation should explicitly define what AI audits entail (47\% -- exemplified by O’Sullivan: “I'm worried about ethics washing [if regulation required ‘audits’], then we would see a whole rush of disingenuous businesses pop into place to be able to fill that role…”), and those who say regulation should provide standards and guidelines, but not specific definitions (53\% -- from O’Neil: “One of the biggest concerns I have right now is that what it means to be audited will be codified in various contexts and then it'll be too narrow.”) 

\subsection{Barriers to Effective AI Auditing}

\begin{quote}
    “[Companies are] there to make money. And the truth is, it's expensive to put constraints of fairness onto their algorithms.” – Interview with Cathy O’Neil, CEO of ORCAA 
\end{quote}

\subsubsection{Lack of auditee buy-in is widespread}
The auditors we surveyed rank ‘lack of buy-in from potential auditees to conduct an audit in the first place’ and ‘cost of conducting an audit’ as the two biggest barriers they face. Our interviewees corroborate this. Several attribute the lack of auditee buy-in to beliefs that performing audits could expose them to legal liability. O’Sullivan (“A lot of existing legal teams and, to some degree, their data science teams as well, are more inclined to believe that they have some protection from legal inquiry if they don't know the degree to which their models are discriminating”), R1 (“[companies] say we don't actually want to investigate our data, because if we do, we're going to find a problem, then we're gonna have to fix it”), and Whittaker (“lawyers would just not allow [internal audits] because a part of liability mitigation is maintaining plausible deniability around harm”) all describe experiencing this directly. Half of our interviewees also explicitly mention cost as a barrier. In terms of unmet needs (Table 3), we find that even when auditors receive sign-off to conduct an audit, their oversight powers remain limited: the majority (65\%) report that auditees will not commit to address problems uncovered by audits, and half report limited power to require changes before deployment (53\%) or to engineering practices (51\%). Most auditors (80\%) have recommended changes to AI systems that were not implemented.

\subsubsection{Barriers differ between first-, second-, and third-party auditors}
\begin{table*}
\centering
\caption{Unmet need as reported by survey respondents\label{unmet_need}}
\begin{small}
\begin{tabular}{cl}
    \toprule
    \textbf{\% Unmet} & \textbf{Description of need} \\
    \midrule
    65\% (28/43) & Accountability: Commitment from auditee to address problems uncovered by audit within set time \\ 
    53\% (23/43) & Enforcement: Ability to require specific changes to AI system before deployment \\ 
    51\% (22/43) & Root cause: Ability to require changes to AI engineering practices\\ 
    42\% (18/43) & Access to harm/incident reports created after deployment \\
    40\% (17/43) & Access to outputs/decisions made by AI system \\ 
    37\% (16/43) & Disclosure: ability to publicly disclose (publish an audit report) \\ 
    37\% (16/43) & Documentation: Ability to create internal audit report \\ 
    35\% (15/43) & Access to individuals who have interacted/will interacted with AI system in the real world \\ 
    35\% (15/43) & Access to technical inputs to AI system \\ 
    23\% (10/43) & Access to individuals involved with building/deploying AI system \\ \bottomrule
\end{tabular}
\end{small}
\Description{A table outlining the areas of unmet need that auditors we surveyed (N=43) faced. 65\% of auditors had unmet need around `commitment from auditees to address problems uncovered by an audit within a set time; 53\% around ability to require specific changes to AI systems before deployment; 51\% around ability to require changes to AI engineering practices; 42\% around access to harm/incident reports created after deployment; 40\% around access to outputs/decisions made by the AI system; 37\% around ability to publicly publish audit reports or create internal audit reports; 35\% around access to individuals who have interacted with the AI system in the real world; 35\% around access to technical inputs to the AI system; and 23\% around access to individuals involved with building/deploying the AI system.} 
\end{table*}

Although all types of auditors face some barriers related to auditee buy-in, some unmet needs differ between first-, second-, and third-party auditors. No first-party auditors report lacking necessary access to data or technical aspects of the audited system, unlike 58\% of second-party and 50\% of third-party auditors. R2, a third-party auditor who reports scraping their own data to conduct audits, explains: “[our] biggest barrier is probably legal. The Computer Fraud and Abuse Act criminalizes terms of service violations, which often occur in automated data collection at scale of publicly available data.” Noting that a federal court in Sandvig v. Barr \cite{ACLUSandvig} ruled that research into algorithmic discrimination does not violate the CFAA, R2 says the issue nevertheless remains: "courts are divided about this, and there's hope based on a recent... ruling that maybe this won't be used to prosecute journalists, but it constantly hangs as a specter over the work we do." Both first- and second-party auditors report being unable to disclose audit results (30\% report ‘Never’ to ‘About half of the time’); while third-party auditors generally do not report this barrier (50\% report that they ‘Always’ disclose). First-party auditors in particular face barriers relating to the corporate structure of their employers. R1 describes grappling with a trade-off between oversight and buy-in from modeling teams: “The safest place for us to live is in risk and compliance. It is also the worst place for us to sit. Because everybody hates working with [them]... no model owner is happy when they get an email from a lawyer.” Some interviewees also discuss issues with company culture more broadly: R3, internal audit lead at a technology company, shares that “[there are] barriers put forward for people from marginalized groups who work in tech where a company can be pro-inclusion, but very much lean in to alienation and exclusion.”

\subsection{Stakeholder Involvement and Real-World Harm}

\begin{quote}
“[We] had meetings with people in civil society, and they would try to distill some of their feedback…[But] they fired the people who [were] against some of these ICE contracts. So, they're not going to [say] `hey, let’s talk to impacted communities and not do this thing that will give us billions of dollars.’ No. Yeah, they don't do that.” – Interview with R4, internal audit lead at a technology company
\end{quote}

\subsubsection{The current state of harm incident reporting is ad hoc}

Of the 46 auditors who answered “Which of the following best describes your approach to AI bias/harm incident reporting?”, 35\% say that they are either not familiar with bias/harm incident reporting or that it is not part of their audit process. Some interviewees say that social media is the only monitored system they are aware of for reporting real-world AI harm. A further 30\% of survey respondents say that while they consider reports of real-world harm, it is not a critical part of their process. Interviewee R1 indicates that even when established, harm reporting systems are not widely used: “we actually already have [harm reporting] systems in place. [It’s] a really good example of just because you have a system in place it doesn't automatically solve the problem.” Others, like R3, feel that harm incident reporting should be standardized: “There really does need to be some sort of standard place to report [algorithmic harms]... We're not there yet. [We’re still] scrappy, searching for what people say and really relying on activists to be active.”

\subsubsection{Less than half of AI auditors say that stakeholder involvement is key, and only 4\% provide documentation of stakeholder involvement}

41\% of auditors who replied to the question (19/46) claim that “involving those who are most at risk of harm” is critical. Liz O'Sullivan, CEO of Parity, put it this way: “The signals that you're getting from data are almost the least interesting thing that you get from these audits because the risk comes in and the harms come in from things that happened in the real world, from sociological [phenomena], which you really can't understand unless you're interviewing your users.” However, only 4\% of respondents (2) provide a link to documentation of audit methodology with evidence of stakeholder involvement. Our methods do not allow us to determine why individuals did not provide more examples. For some, it may mean that they are not examining specific instances of real-world harm in their audits, even when they say doing so is critical. For others, it may mean they are not at liberty to share examples due to client confidentiality considerations. Some interviewees also describe the challenges associated with stakeholder involvement, including translating stakeholder interviews into actionable results, like R1: “I have seen lots of feel-good moments in design thinking rooms and lots of sticky notes put places, and maybe a couple of product tweaks, but fundamentally…I've never seen any of that translate into model and into code and into data.”		
\section{Discussion and Recommendations\label{discussion}}

\begin{quote}
   “I'm not seeing [regulatory] overstep right now. I'm just seeing wild, wild west. Just anybody doing whatever they want.” – Interview with R4, internal audit lead at a technology company
\end{quote}

\subsection{Discussion}

Overall, three major themes emerged from our research. First, we find that \textit{the algorithmic audit ecosystem, while nascent, is growing rapidly}. Based on the number of individuals (N = 438) and organizations (N = 189) we identified as involved in AI audits, we expect that the need to establish standards and regulatory oversight to reduce discrepancies between auditors’ desired best practices and reality will only become more pressing. Second, \textit{there is a consensus among practitioners that current regulation is lacking, as well as agreement about some areas that require mandates}. Auditors and non-auditors alike overwhelmingly agree that AI audits should be mandated (95\%), and that the results of these audits, in part or in full, should be disclosed (82\%). Auditors also near-universally report that their largest barriers are lack of buy-in from auditees to conduct audits in the first place, and that even when they have buy-in, they have limited enforcement capabilities. Finally, we find a \textit{mismatch between what auditors say is important to conducting audits and what they actually do}. Auditors often want to disclose their results and methods, but are restricted by nondisclosure agreements. Many express interest in intersectional analysis (65\%), but worry that by gathering the demographic data necessary to demonstrate disparate impacts, they may run afoul of anti-discrimination law. Similarly, although many auditors consider analysis of real-world harm (65\%) and inclusion of stakeholders who may be directly harmed (41\%) to be important in theory, they rarely put this into practice. 

\subsection{Policy Recommendations}
Based on the findings outlined above, we propose six policy recommendations that we believe should be prioritized: 1) require both owners and operators of AI systems to engage in independent audits against clearly defined standards; 2) require that individuals be notified when they are subject to algorithmic decision-making; 3) mandate disclosure of key components of audit findings for peer review; 4) consider real-world harm in the audit process, including through standardized harm incident reporting and response mechanisms; 5) directly involve the stakeholders most likely to be harmed by AI systems in the AI audit process; and 6) formalize evaluation and, potentially, accreditation of AI auditors. The first four have broad support from auditors and non-auditors alike; the last two are more controversial among audit practitioners. Each recommendation has implications for lawmakers and regulators as they craft bills and regulatory mechanisms to govern AI systems, for private companies as they develop internal policies, and for standards-setting bodies and professional associations as they organize consensus around standards and best practices. Our recommendations are intended as core ideas, rooted in lessons from the existing community of practice. Future work will be required to link these to existing policy, to develop practical recommendations for implementation, and to develop additional domain-specific standards.

\subsubsection{Require audits for AI system owners and operators}
AI auditors are in near-universal agreement (95\%) that audits should be a required element of owning and operating AI systems. Across the board, auditors say that their most important regulatory priority is to establish legislation that requires companies to engage in AI auditing. They also rank “lack of buy-in from potential auditees to conduct an audit in the first place” as their biggest barrier. Rather than let companies choose whether, how, and when to conduct audits, policymakers should enact legislation and regulatory requirements for AI vendors and operators to submit to audits. However, many open questions remain. For example, it is unclear whether audit mandates should only apply to a restricted set of applications in high-stakes contexts, or whether all AI systems should be subject to audits. The pending legislation is split here too -- while the EU's AI Act and DSA would only require audits of `high-stakes' or `very large' sytems, NYC's recently passed legislation will require audits of all ADS hiring tools \cite{EUAIAct, EUDSA, NYCAudit}.
The auditors we spoke to are also split over how tightly audits should be defined by law. Some worry that companies could take advantage of overly broad language to conduct cursory checks, while others worry that too-specific language would lead to overly narrow assessments. Finally, over half (65\%) of the auditors we surveyed indicate that they receive neither commitment from auditees to address problems uncovered by audits nor the power to require specific changes to an AI product or to engineering practices. This indicates that regulators should also develop compliance mechanisms to ensure that audits lead to real change.

\subsubsection{Notify individuals when they are subject to algorithmic decision-making}
Mandatory notification when individuals are subject to algorithmic decision-making is a widely supported regulatory intervention among our study respondents (for example, O'Sullivan: "I'd love to see a requirement that people understand when AI is touching their lives"). Notification is a first step that enables individuals to request additional information about, and potentially contest, decisions made by AI systems. In the public sector, notification also provides a basis for communities to demand oversight, and to weigh in on whether AI systems should be deployed in the first place. We therefore believe that policymakers should require notification, opt-out, and appeal. Of course, policy alone does not automatically change practice. For example, GDPR includes a duty to notify, but this is not always respected \cite{GDPR, GDPRDeloitte, GDPRNYT, GDPRWhatsApp}. Accountability will require awareness, ongoing monitoring, and, when necessary, litigation; audit standards should include evaluation of whether notification, opt-out, and functional decision appeal mechanisms are in place.

\subsubsection{Mandate disclosure of key audit results to external stakeholders}
The vast majority of AI auditors (82\%) view disclosure of audit results as a necessary part of the audit process. However, most do not yet disclose audit methods or findings themselves (only seven provided links to any documentation; just four linked to audit results), typically citing client confidentiality concerns. At the same time, the AI auditors that survey respondents describe as “best in class” are all authors of public, peer-reviewed audits. We believe that this speaks to the importance of disclosure. Third-party auditors have unique incentives to disclose, as R2 (a third-party auditor) emphasizes: “We provide the methodology, plus all the data and code that we used, to the target of the [audit]...we think [they have] a really big incentive to find something wrong and we want them to find that thing wrong because then it will help us make our findings stronger.” In response to the issue raised by O’Neil (“I can't demand [disclosure]. [Clients] simply won't hire me.”), we note that if disclosure of key audit results and methods is legally mandated, then AI owners and operators will no longer be able to demand complete confidentiality from second-party audit contractors. For this reason, we believe that legislation requiring disclosure of key audit results and methods is a high priority. That being said, the mandated degree of disclosure must be considered carefully, and may benefit from domain-specific legislation. If every detail of all audits were to become public, this information could be used in some cases by competitors and/or malicious actors. Should audits be shared publicly (like food safety audits or health inspections \cite{HealthInspections}), or logged in a database and made accessible only via request by vetted actors (like the International Air Transport Association's Operational Safety Audits \cite{IATAIOSA})? Should disclosure cover all details of the audited system, audit process, and audit results, or should auditors be required to report only an overview of methods and key findings? Should the degree of disclosure be tied to the level of risk of harm from the AI system? Adequate disclosure regulation requires working through these details.

\subsubsection{Require harm incident reporting and other mechanisms for real-world harm discovery}
Many auditors say that, while their organization or their clients’ organizations have systems in place for users to report experiences of harm, they believe that those systems are ineffective (R1: “just because you have a system in place it doesn't automatically solve the problem.”) or ad hoc (Polli: “You can absolutely send us feedback. But we don't solicit information.“). Surveyed auditors report being more likely to assess the quantitative aspects of real-world harm than the structural or qualitative aspects; less than half report that they assess whether mechanisms for whistleblower protection (24\%) or appeals processes (46\%) are sufficient, examine the environmental impact of the AI system (37\%), or determine whether the AI system uses data sources that rely on unfair labor practices (48\%). We believe that it is important to establish policies that require the consideration of real-world harm in audit processes, including standardized methods for harm incident reporting and mechanisms to ensure that AI owners and operators respond to harm reports. Lawmakers and regulators should require harm incident reporting by AI vendors and operators, and establish incident reporting and response standards, mechanisms, and databases.

\subsubsection{Prioritize stakeholder involvement by affected communities}
Less than half (41\%) of auditors say that stakeholder involvement is currently a critical part of their audit process, and of those, just 30\% say that they consider real-world harm to stakeholders (only two provide linked examples). Too often, stakeholder engagement does not actually inform AI product design decisions or audits. It should be a priority for regulators to ensure that audits include affected stakeholders, and for organizations to establish internal policy that promotes direct involvement of the stakeholders most likely to be harmed by AI systems. Although the benefits of broader stakeholder participation are well understood, executing on this vision is challenging and expensive \cite{DesignJustice}. As Whittaker mentions, “it's messy and there isn't a rubric. But it isn't just holding a town hall and being like `speak.' […] Oftentimes it's taking the train down there, and sitting with folks and getting a feel for the thing.” Solutions should be informed by the existing field of participatory design and by the growing community of design justice practitioners,\footnote{For a recent in-depth review of participatory design of socio-technical systems, see Costanza-Chock \cite{DesignJustice}.} and should be supported by field-wide investment in strategies to meaningfully engage community partners and support community-led processes for algorithmic accountability.

\subsubsection{Formalize evaluation and consider accreditation}
Formal accreditation of AI auditors is not particularly popular among those we surveyed (when asked about regulatory priorities, 75 respondents ranked this proposal last). And there are many legitimate concerns about AI auditor accreditation -- accreditation should not become a `rubber stamp' process, nor should it lock out independent researchers, civil society organizations, investigative journalists, and community advocacy organizations who may have extensive lived experience of and high motivation to expose harmful AI systems. At the same time, many AI auditors express concern that their peers do not engage in rigorous audit practices, or that companies would perform cursory audits of their own algorithms if given the chance. As O'Neil puts it: ``My fear is that some of the people who are doing AI ethics are just naive. I think they can be technically quite competent, but if they don't recognize that [narrow audits are] something that could amount to gaming, then they could be taken in." In other industries, accreditation is typically provided by an accreditation body or bodies. For AI auditing, this might take the form of a professional organization for AI auditors, a public agency, and/or an international standards-setting body to provide accreditation guidance. Finally, we note that many AI auditors currently work with customized frameworks and ad-hoc tools, and that there are not many resources available to support their training and methodological development. Whether or not auditors are formally accredited, the field would be greatly strengthened by deep investment in the development of educational curricula and infrastructure, improved tools and technical frameworks, and mentorship to grow and diversify the talent pool of capable auditors.

\section{Conclusion}

If auditing is to evolve into a key mechanism for algorithmic accountability, it is important to understand and shape the emerging ecosystem of AI auditors. In this paper, we provide an overview of the current state of the algorithmic audit field through a series of interviews with leading audit practitioners and a survey aimed broadly at first-, second-, and third-party auditors; advocates; regulators; and researchers. We find that the AI audit ecosystem is growing rapidly and that practitioners in the field overwhelmingly believe that current regulation is lacking, underscoring the need to establish clear regulatory requirements and standards. We identify significant areas of consensus around the need for mandatory audits, notification, and disclosure of key audit findings; we also outline areas of debate about the details of emerging regulation and standards. Additionally, we find a mismatch between what auditors say is important and what they are currently able to accomplish in practice. To address these challenges, we outline recommendations for regulatory and organizational policies that we believe will enable auditors to play a more meaningful role in reducing algorithmic harm. These include: 1) mandatory independent AI audits against clearly defined standards, applicable to both AI product owners and operators; 2) required notification to individuals when they are subject to algorithmic decision-making systems; 3) mandated disclosure of key components of audit findings for peer review; 4) consideration of real-world harm in the audit process, including standardized harm incident reporting and response mechanisms; 5) stakeholder participation in audits, in particular by communities most likely to experience harm from the system, product, or tool that is being audited; and 6) a formal system for evaluation and, potentially, accreditation of AI auditors.

\begin{acks}
We'd like to thank the many research collaborators and participants who made this project possible.
\end{acks}

\bibliographystyle{ACM-Reference-Format}
\bibliography{references}

\appendix

\section{Methods}

\subsection{Field Scan}
We first conducted a field scan that yielded a list of 438 individuals from 189 organizations that were involved, to some degree, in AI auditing. This group included first-, second-, and third-party auditors as well as members of civil society and advocacy organizations, academic researchers, regulators involved with AI audit legislation and policy, and those otherwise involved in work related to audits. Within that group, we identified ten (10) leaders in the field for semi-structured interviews. Based on our understanding of the field and lessons from our interviews, we developed a survey that we shared with contacts at all 189 organizations we identified in the field scan. 

\subsection{Interviews\label{methods-interviews}}
We conducted ten semi-structured interviews with individuals who we identified as leaders in the field of AI auditing. We gave interviewees the option to remain anonymous; those who did not exercise this option were: Frida Polli, CEO of pymetrics (joined by Alex Vaughan, Chief Science Officer); Cathy O’Neil, CEO of O’Neil Risk Consulting and Algorithmic Auditing (ORCAA); Liz O’Sullivan, CEO of Parity; Reid Blackman, CEO of Virtue; Michelle Lee, AI Ethics Lead at Deloitte UK; and Meredith Whittaker, Faculty Director of AI Now. Four interviewees chose to remain anonymous; they included three first-party auditors (one at a social media company and two at technology companies) and one third-party auditor. Throughout, we attribute quotes and insights to the identified interviewees and denote anonymous interviewees with R1, R2, R3, and R4. Each interview lasted 45-60 minutes and was conducted and recorded via Zoom. All interviewees provided informed consent prior to the beginning of the interview. In each interview, we sought to understand (1) the interviewee’s organization and their role within it, (2) the interviewee’s perspective on the industry, and (3) if and how the interviewee and/or their organization navigates potential and/or actual harm. Each interview followed a semi-structured interview guide with three sections, each intended to address one of our interview objectives. 

The sections contain lead questions (e.g. “What methods and tools do you currently use to audit AI systems”) that were asked to all interviewees as well as follow-on questions (e.g. “Can you talk about any specific standards that you audit against?”) which were asked to interviewees depending on their relevance to the rest of the interview. The section on interviewee’s industry perspective included questions on the methods and tools currently used to audit AI systems, perceived emerging standards and best practices, experienced barriers to effective AI auditing, and desired regulatory oversight. Throughout this section, we first asked open-ended questions (e.g. “What are some of the biggest barriers to effective AI auditing that you have experienced?”) and followed up with more specific prompts (e.g. “Do any of these common barriers apply to your team's experience?...”) if necessary. To understand the interviewee’s approach to real-world harm, our team asked questions related to the presence of stakeholder involvement in the audit process, performance of harm investigation at any point during the AI lifecycle, and the review of harm reporting for deployed systems. Although we began with open-ended questions in this section as well (e.g. “Do your audits consider what system (if any) is in place for people to report experiences of harm?”), we provided examples if interviewees were not familiar with terms like `harm reporting systems’ (“For example in cybersecurity, best practice includes incident reports, verification, and escalation – do you implement anything like that for algorithmic harms?”), `stakeholder involvement’ (“For example, do you conduct interviews, focus groups, or workshops that involve organizations representing marginalized communities?”), and `potential or actualized real-world harm’ (“For example, people might be harmed during development, such as in nonconsensual data use or bad labor practices in labeling; or after deployment, such as if they are incorrectly assessed or classified by an AI system.”) 

\subsection{Survey}
Following the completion of the interview phase, and to further explore the research questions outlined above, our research team developed an online survey. The survey was open from 9/23/2021 to 10/29/2021. Initially the survey was sent directly to individuals from our list of 438 individuals across 189 organizations via email, LinkedIn, and Twitter direct messages. Later, the survey link was released publicly via social media (LinkedIn and Twitter) and shared over the Algorithmic Justice League's newsletter mailing list. Overall, we received 152 individual survey responses, with 57 of those from the invited list and 95 from public outreach. Our survey included respondents from six continents and twenty-five countries. 59\% of all respondents (90 individuals) were from the United States.

Survey respondents were asked to identify their relationship to AI auditing. If an individual selected that they had personally conducted audits, they were directed to questions that pertained specifically to AI audits (e.g., "which of the following best describes your AI auditing process?"). Overall, 37\% (56) indicated they had personally worked on an audit of an AI system whereas 63\% (96) identified themselves as non-auditors. Of those who identified themselves as auditors, respondents came from five continents and twelve countries with 61\% from the United States (34 individuals out of 56).
\section{AI Audit Field Scan}

\href{https://docs.google.com/spreadsheets/d/1NR-29waYpARervQYonuLdl4_G5Ei7E3cmq_Z-dMa7-M/edit?usp=sharing}{\underline{Linked}} is a Google sheets spreadsheet of organizations and individuals involved to some extent in algorithmic audits.

\section{Interview Instrument}

\href{https://drive.google.com/file/d/1nSVgceVzzTU7Q0nm_Q-CNrDXj6btweFf/view?usp=sharing}{\underline{Linked}} is a PDF version of our interview guide. The interview guide was sent to interviewees once an interview was scheduled. In the interview guide, we provided an introduction, directions, a research interview consent agreement, and questions.

\section{Survey Instrument \label{survey_instrument}}

\href{https://drive.google.com/file/d/1Oial_0sRAyXrNWu2iuj_OZB5WBCex00i/view?usp=sharing}{\underline{Linked}} is a PDF version of our survey questions. Note that this PDF also contains the full list of AI audit toolkits mentioned in Section \ref{tools}.

\end{document}